\documentclass[useAMS,usenatbib]{mn2e}

\usepackage{graphicx}

\usepackage{natbib}
\newcommand{\be}{\begin{displaymath}}
\newcommand{\ee}{\end{displaymath}}
\newcommand{\bea}{\begin{eqnarray}}
\newcommand{\eea}{\end{eqnarray}}
\newcommand\msol{M_{\odot}}

\newcommand{\msun}{\ensuremath{\, M_\odot}}

\newcommand{\abb}[1]{Fig.\,\ref{#1}}

\title[Super-massive stars in globular clusters]{Super-massive stars as a source of abundance anomalies of proton-capture elements in globular clusters }

\author[P. A. Denissenkov and F. D. A. Hartwick]{P. A.
Denissenkov$^{1,2}$\thanks{E-mail: pavelden@uvic.ca.} and F. D. A. Hartwick$^{1}$\\
$^{1}$Department of Physics \& Astronomy, University of Victoria,
       P.O.~Box 3055, Victoria, B.C., V8W~3P6, Canada\\
$^{2}$The Joint Institute for Nuclear Astrophysics, Notre Dame, IN 46556, USA}
\begin{document}

\date{Accepted 2013 September 22. Received 2013 September 6; in original form 2013 August 19}

\pagerange{\pageref{firstpage}--\pageref{lastpage}} \pubyear{2013}

\maketitle

\label{firstpage}

\begin{abstract}
We propose that the abundance anomalies of proton-capture elements in globular clusters, such as the C-N, Na-O, Mg-Al and Na-F
anti-correlations, were produced by super-massive stars with $M\sim 10^4 \msun$. 
Such stars could form in the runaway collisions of massive stars that sank to the cluster centre as a result of
dynamical friction, or via the direct monolithic collapse of the low-metallicity gas cloud from which the cluster formed.
To explain the observed abundance anomalies, we assume that the super-massive stars had lost significant parts of
their initial masses when only a small mass fraction of hydrogen, $\Delta X\sim 0.15$, was transformed into helium.
We speculate that the required mass loss might be caused by the super-Eddington radiation continuum-driven stellar wind or by
the diffusive mode of the Jeans instability.

\end{abstract}


\section{Introduction}
\label{sec:intro}

The history of globular-cluster abundance anomalies dates back to the 1970's \citep[e.g.][]{kraft:79}.
A recent comprehensive review of the problem is given by \cite{gratton:12}.
In this review a compilation of the Na-O anti-correlations
for a number of globular clusters (GCs) is shown. Other anti-correlations are also observed
(e.g. C-N, Mg-Al, and Na-F) but the one involving Na and O has come to practically
define the GC multiple population problem. The combinations of chemical elements
in these anti-correlations unambiguously indicate that they resulted from hydrogen burning
in the CNO, NeNa, and MgAl cycles \citep{denisenkov:90,langer:95}. The direct evidence of the He abundance variations
between red giants in GCs \citep[e.g.][]{pasquini:11,dupree:11} supports this conclusion.
However, the fact that the similar abundance anomalies are found not only in red giants but also in main sequence (MS)
turn-off and early subgiant stars \citep[e.g.][]{gratton:01} implies that they were produced by 
the extinct first generation of GC stars.

Presently there are two main proposed sources for polluting the original
gas in GCs. The first is fast rotating massive stars \citep{decressin:07} and
the second is hot-bottom burning (HBB) in intermediate-mass and massive AGB stars \citep{dantona:83,ventura:13}. 
%In the first case, the mass lost from the rotating stars is assumed to form the
%second generation of stars, while in the second case it is mass lost from the
%first generation AGB stars. 
Examples of how the AGB stars might solve the
GC multiple population problem are given in the works of \cite{bekki:07}, \cite{dercole:10}, and
\cite{herwig:12}.
From the nucleosynthesis point of view, potential problems with these sources are 
that H burning in the MS stars with $20 \msun\leq M\leq 120 \msun$ 
considered by \cite{decressin:07} occurs at central temperatures $T_\mathrm{c} \la 60\times 10^6$ K 
that are not sufficient to transform $^{24}$Mg into Al, as is often seen in GC stars, while
the physical conditions for the required O depletion during HBB in the AGB stars lead to an even stronger Mg destruction
\citep{denissenkov:03} or to a deficit of Na \citep{ventura:13}, neither of which has been observed.

In this Letter, we propose super-massive stars (hereafter, SMSs) with $M\sim 10^4 \msun$ as a main source of 
the abundance anomalies of proton-capture elements in GCs. Our solution of
the problem has been motivated by (1) the accumulating evidence of the presence of intermediate-mass black holes (IMBHs) in
GCs \citep[e.g.][]{lutzgendorf:13}, that could be remnants of SMSs, 
(2) the results of numerical simulations of runaway massive star collisions     
that explain the formation of SMSs in the cores of young dense star clusters \citep{portegies:99,portegies:04}, 
and (3) the astonishing resemblance of theoretical anti-correlations between the p-capture element abundances obtained 
with our SMS models to the ones observed in GCs.

\section{The Formation of SMSs in GCs}

\cite{portegies:04} performed numerical simulations of the evolution and motion of stars in two young clusters,
MGG\,9 and MGG\,11, located near the centre of
the starburst galaxy M\,82. To explain the presence of a luminous X-ray source, presumably an IMBH with
a mass of at least $350 \msun$, in MGG\,11 and the absence of such an object in MGG\,9, 
they estimated masses and projected half-light radii of
the two clusters and came to the conclusion that only MGG\,11 was dense enough for massive zero-age MS stars to be able to sink
to its centre, as a result of dynamical friction, before exploding as supernovae. After reaching the centre, 
the massive stars are predicted to undergo multiple collisions with each other in a runaway process that eventually
leads to the formation of an SMS \citep{portegies:99}. \cite{portegies:04} reported that for the cluster MGG\,11,
whose total estimated mass was $(3.5\pm 0.7)\times 10^5 \msun$, the mass of such
a runaway merger product should be $\sim 10^3 \msun$. Later,
\cite{portegies:06} investigated the possibility that the IMBH remnants of runaway SMSs in young dense star clusters that formed
within the inner 100\,pc of the Milky Way could contribute to the growth of the supermassive black hole in the Galactic centre.
They used the following fitting formula for the mass of the runaway SMS:
\bea
M_\mathrm{SMS}\sim 0.01 M_\mathrm{tot} \left(1+\frac{t_\mathrm{rh}}{100\ \mbox{Myr}}\right)^{-1/2},
\label{eq:MSMS}
\eea
where $M_\mathrm{tot}$ is the total cluster mass, and $t_\mathrm{rh}$ is the dynamical relaxation time at the mean density of
the inner half of the cluster's mass. Equation (\ref{eq:MSMS}) was calibrated by N-body simulations for a Salpeter-like IMF.
For the range of initial masses of GCs that formed nearly 12\,Gyr ago and survived until the present day,
$4\times 10^5 \msun\la M_\mathrm{tot}\la 6\times 10^6 \msun$ \citep{gnedin:08}, equation (\ref{eq:MSMS})
gives $4\times 10^3 \msun\la M_\mathrm{SMS}\la 6\times 10^4 \msun$ for $t_\mathrm{rh} \la 100$ Myr.
%However, given that many of the present-day GCs have $t_\mathrm{rh}\sim 1000$ Myr \citep{harris:10}, 
%the estimated $M_\mathrm{SMS}$ values should probably be reduced by a third. 
We realize that our extrapolation of equation (\ref{eq:MSMS}) beyond the limit of $M_\mathrm{tot}$, for which it was originally
fitted, may overestimate the upper limit of $M_\mathrm{SMS}$. For comparison,
the mass of a collapsing core in a young dense star cluster calculated by \cite{gurkan:04} is
$M_\mathrm{cc}\sim 0.001$\,--\,$0.002 M_\mathrm{tot}$ and it
weakly depends on initial conditions. \cite{freitag:06} modeled runaway collisions between stars in such collapsed
cores and found that the stellar merger product grows rapidly to $M_\mathrm{SMS}\ga 1000 M_\odot$, however, they could not 
accurately predict its final mass.
It is interesting that the masses of IMBHs measured in GCs seem to correlate with their $M_\mathrm{tot}$
\citep{kruijssen:13}, which is in a qualitative agreement with formula (\ref{eq:MSMS}), provided that
$M_\mathrm{IMBH}\propto M_\mathrm{SMS}$. Besides, three GCs probably host IMBHs with the masses in excess of $10^4 M_\odot$.
These arguments give us grounds to explore the hypothesis that the GC abundance anomalies were produced by SMSs with
$M_\mathrm{SMS}\sim 10^4 M_\odot$.

\section{The Core Hydrogen Burning in SMSs}

To model the MS evolution of SMSs, we have used the MESA {\tt star} module with its default opacities,
EOS, and reaction rates \citep[see \S\,4 in][]{paxton:11}. 
Our adopted nuclear network includes 31 isotopes from $^1$H to $^{28}$Si
coupled by 60 reactions of the pp chains, CNO, NeNa, and MgAl cycles.
Because the SMSs are radiation pressure dominated objects with
super-Eddington luminosities, we have used a new MLT++ treatment of convection recommended in MESA for such cases 
\citep[see its discussion in \S\,7.2 in][]{paxton:13}. For the initial chemical composition,
we have assumed the helium and heavy-element mass fractions $Y=0.25$ and $Z=5\times 10^{-4}$, i.e.
[Fe/H]\,$\approx -1.6$, with the $\alpha$-element
abundance enhancement $[\alpha/\mbox{Fe}] = +0.4$ and solar relative abundances of the Mg isotopes. 

We postulate that SMSs in young GCs continue to burn H in their cores only until the central He abundance increases 
by $\Delta Y = 0.15$ which approximately corresponds 
to the largest difference in $Y$ between Na-poor and Na-rich sub-populations of GC stars \citep[e.g.][]{pasquini:11,king:12}.
After that, the SMSs are assumed to lose the greater parts of their masses, presumably as a result of various instabilities and 
stellar winds. The remaining parts of the SMSs eventually collapse to 
directly form IMBHs \citep{heger:03}.
The mass lost from the SMSs is mixed with the original gas in different proportions, and the second generation of GC stars is formed 
from this mixture.

We have found that the best agreement with observations is obtained for SMSs with masses close to the upper limit
estimated with formula (\ref{eq:MSMS}). In \abb{fig:f1} and \abb{fig:f2}, the observed correlations
between p-capture isotopes for MS turn-off, subgiant and red giant stars 
in the GCs $\omega$\,Cen, M\,13, M\,71, M4, NGC\,6397, NGC\,6752, and 47\,Tuc
are compared with their corresponding
theoretical counterparts from our SMS models calculated for $M_\mathrm{SMS} = 2\times 10^{4} \msun$, $3\times 10^4 \msun$, and
$4\times 10^4 \msun$. The MS stars with these masses are fully convective, their super-Eddington luminosities lie
between $\sim 10^9 L_\odot$ and $\sim 10^{10} L_\odot$, and $T_\mathrm{eff}\sim 10^5$ K.
The symbols on the theoretical curves mark abundances in mixtures composed of an $f$ part of the original 
gas and a $(1-f)$ part of the material from our SMS models at the moment when $Y = 0.40$. The parameter $f$ decreases
from 1 to 0 with the step $\Delta f = 0.1$ starting from the following abundance ratios measured in
the GC stars that are believed to belong to the first generation: $^{26}\mbox{Mg}/\mbox{Mg} = 11\%$, 
$^{25}\mbox{Mg}/\mbox{Mg} = 10\%$, $^{24}\mbox{Mg}/\mbox{Mg} = 79\%$, [Al/Fe]\,$=0$, [Na/Fe]\,$=-0.4$, [Mg/Fe]\,$=+0.4$,
[O/Fe]\,$=+0.4$, [F/Fe]\,$=0$, [N/Fe]\,$=0$, and [C/Fe]\,$=0$.

The agreement between the observations and theory is
very good, especially in \abb{fig:f1}, including the correlations of the Mg isotopic ratios with [Al/Fe] 
recently complemented by \cite{dacosta:13}. From the nucleosynthesis point of view, this result is not
surprising because \cite{denissenkov:98} demonstrated that the GC abundance 
anomalies could be the products of H burning at $T\approx 74\times 10^6$ K, that had consumed only a few percent of 
initial hydrogen, mixed with the original gas. However, the main stellar source in which
these conditions can be realized has remained unknown. We think that the SMSs with masses $\sim 10^4 \msun$
could be such a source because our models do have the required central temperatures 
$74\times 10^6\,\mbox{K}\la T_\mathrm{c}\la 78\times 10^6\,\mbox{K}$. Besides, the plausible mechanism of their formation 
via the runaway star merger in
young dense star clusters that are similar to young GCs \citep{portegies:04} as well as the possible presence of IMBHs with masses 
up to $4.7\times 10^4 \msun$ in the present-day GCs \citep{lutzgendorf:13} support our hypothesis.
Our calculations also show that MS stars with $M\sim 10^3 M_\odot$ only reach $T_\mathrm{c}\sim 60\times 10^6$ K
which is not high enough for them to be a main source of the abundance anomalies of p-capture elements in GCs, 
as was earlier demonstrated by \cite{sills:10}.

%-----------------------------------------------------------------
\begin{figure}
%\epsfxsize=10cm
%\epsffile[40 150 380 650] {f1.eps}
\includegraphics[scale = 0.5, bb = 85 140 500 700]{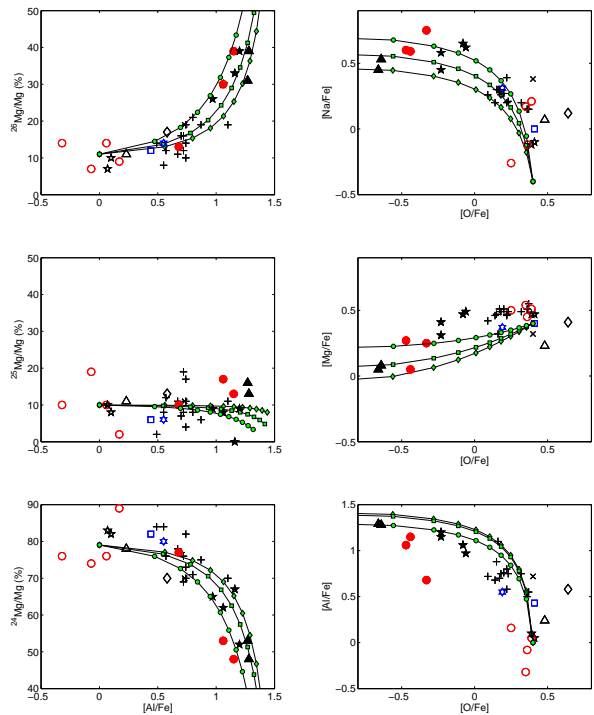}
\caption{The correlations and anti-correlations between the proton-capture isotopes for stars in the GCs
         $\omega$\,Cen, M\,13, M\,71, and M\,4 taken from the work of \protect\cite{dacosta:13} (we used their
         original symbols; for data sources, see the caption to Fig. 4 in the cited paper) are compared
         with our model abundances (green symbols connected by curves) obtained from mixtures of the GC original 
         gas and matter lost by the $2\times 10^4 \msun$ (circles), $3\times 10^4 \msun$ (squares), and
         $4\times 10^4 \msun$ (diamonds) SMSs. The fraction of the original gas in the mixture decreases 
         from 1 to 0 with the step 0.1, starting from the assumed initial abundances (see text).
         }
\label{fig:f1}
\end{figure}
%-----------------------------------------------------------------

%-----------------------------------------------------------------
\begin{figure}
%\epsfxsize=10cm
%\epsffile[40 150 380 650] {f2.eps}
%\epsffile[60 125 480 695] {f2.eps}
\includegraphics[scale = 0.5, bb = 70 160 400 670]{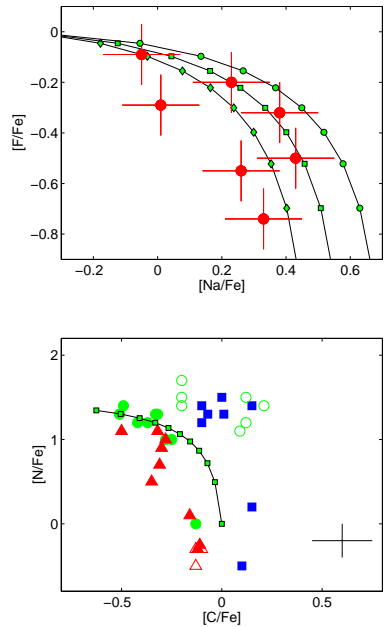}
\caption{Upper panel: same as in \abb{fig:f1}, but for the Na-F anti-correlation for M\,4 red giants from
         the work of \protect\cite{smith:05}. Lower panel: same as in \abb{fig:f1}, but for the C-N anti-correlation
         reported by \protect\cite{carretta:05} for the slightly evolved stars in the GCs GC\,6397 (blue squares), 
         NGC\,6752 (green circles), and 47\,Tuc (red triangles). Open symbols are dwarfs and filled symbols
         are subgiants. For the C-N case, the three SMS models predict the same results. 
         Crosses in the panels show the errorbars.
	 }
\label{fig:f2}
\end{figure}
%-----------------------------------------------------------------

\section{Mass Loss From SMSs}

Although there have been many attempts to estimate mass loss rates for very massive and super-massive stars and
include them in stellar evolution computations
\citep[e.g.][]{owocki:04,smith:06,yungelson:08,dotan:12}, they still remain very uncertain, especially for objects with
super-Eddington luminosities. The most relevant results presented in these works can be summarized as follows
\citep{dotan:12}. The radiative luminosity stays below its Eddington limit 
in deep interiors of SMSs, where convection is very 
efficient and able to carry the super-Eddington energy flux. Closer to the atmosphere, where $L$ approaches $L_\mathrm{Edd}$, 
various dynamical instabilities create spatial inhomogeneities
\citep[``a porous atmosphere'',][]{owocki:04}. This reduces
the effective opacity, which increases $L_\mathrm{Edd}$ above its classical value. Finally, at a larger radius,
the clumps become optically thin, allowing the radiative flux to accelerate them against the gravity force.
This produces a continuum-driven stellar wind. It is believed that such super-Eddington winds, driven by the continuum
radiation pressure, might be responsible for the extremely high mass loss rates, $\dot{M}\sim 0.01$\,--\,$0.5 \msun/\mbox{yr}$,
in $\eta$ Car and other luminous blue variable (LBV) stars \citep{smith:06}. In our SMS models, such a strong
stellar wind could disperse a significant amount of mass by the time that $Y$ has increased by $\Delta Y = 0.15$, which takes
$\sim 10^5$ years.

\cite{thompson:08} has proposed that massive stars in which the radiation pressure dominates over the gas pressure may be locally
unstable with respect to a diffusive mode of the Jeans instability. The growth timescale of this mode is the Kelvin-Helmholtz time
\bea
t_\mathrm{KH}\sim \frac{3\kappa c_\mathrm{s,r}^2}{4\pi Gc},
\label{eq:tkh}
\eea
where $c_\mathrm{s,r}^2 = (4P_\mathrm{r}/3\rho)$ is the square of the adiabatic sound speed for the radiation pressure
$P_\mathrm{r}$ and density $\rho$, and
$\kappa$ is the Thomson electron scattering opacity. $t_\mathrm{KH}$ is independent of spatial scale,
therefore the diffusive Jeans instability may lead to fragmentation of SMSs on this timescale. 
In \abb{fig:f3}, the solid blue curve shows
a profile of $t_\mathrm{KH}$ in our $2\times 10^4 \msun$ model when $Y=0.40$, while the dashed red line
gives its corresponding age. We see that it is the layers closest to the surface
that may become unstable and begin falling apart first.
Then the deeper layers may get involved in the fragmentation. Eventually, only the central core may be left intact
that will form an IMBH. Note that a reduction of $\kappa$ in equation (\ref{eq:tkh}) caused by the porosity of
the super-Eddington layers should decrease $t_\mathrm{KH}$.

%-----------------------------------------------------------------
\begin{figure}
%\epsfxsize=10cm
%\epsffile[40 150 380 650] {f3.eps}
%\epsffile[60 125 480 695] {f3.eps}
\includegraphics[scale = 0.5, bb = 70 230 400 580]{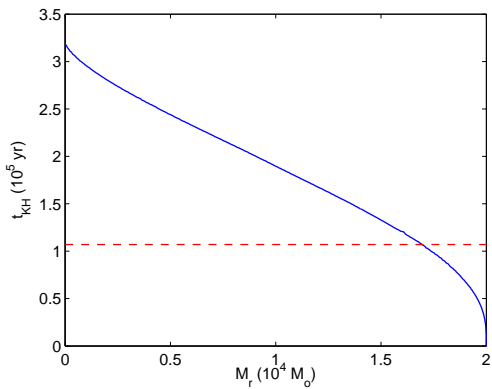}
\caption{Solid blue curve is the profile of the growth timescale for the diffusive Jeans instability
         from equation (\ref{eq:tkh}) in our $2\times 10^4 \msun$ SMS model at the age shown by the red dashed
         curve, by which the helium mass fraction has reached the value $Y=0.40$.
	 }
\label{fig:f3}
\end{figure}
%-----------------------------------------------------------------

\section{Discussion and Conclusion}

The main goal of this Letter is to show that the incomplete core H burning 
in the SMSs with $M_\mathrm{SMS}\sim 10^4 M_\odot$ 
(neither $\sim 10^3 M_\odot$ nor $\ga 10^5 M_\odot$) reproduces the observed abundance anomalies of p-capture elements in GCs
much better than any other mode of stellar nucleosynthesis proposed so far.
Our hypothesis explains very well
the C-N, Na-O, Na-F, $^{24}$Mg-Al anti-correlations and the $^{26}$Mg-Al correlation
(\abb{fig:f1} and \abb{fig:f2}), and the star-to-star variations of the He abundance, $\Delta Y\sim 0.15$, 
in the presently observed GC stars. We realize that, like the fast rotating massive stars, the SMSs destroy lithium, which can be a problem
for explaining the Na-Li anti-correlation and O-Li correlation found in some GCs \citep{dantona:12}. However, given that Li is a fragile element
that survives only in thin surface layers of low-mass MS stars, where its abundance is reduced by atomic diffusion counteracted by additional mixing of
unknown physical origin \citep[e.g.][]{gruyters:13}, we do not think that the Li data can be used as a sole argument against polluting sources involving
massive and super-massive stars.

We have found that, on the MS, the SMSs with $M = 2\times 10^4 \msun$, $3\times 10^4 \msun$, and $4\times 10^4 \msun$ have
the central temperatures close to $T_\mathrm{c}\approx 74\times 10^6$ K that, as was shown by
the ``black box'' model of \cite{denissenkov:98}, is required to reproduce the GC abundance anomalies.
However, like in the ``black box'' model, we have to postulate that only a small mass fraction of hydrogen,
$\Delta X\sim 0.15$, was transformed into He in the SMSs before they lost significant parts of 
their initial masses.
We speculate that it is the super-Eddington radiation continuum-driven wind \citep{dotan:12} or the diffusive mode of 
the Jeans instability \citep{thompson:08} that might result in the required mass loss.

As regards the formation of the SMSs in young GCs, we consider
the runaway collisions of massive stars that sink to the cluster centre through dynamical friction 
\citep{portegies:04,gurkan:04,freitag:06} as a promising mechanism.
It could explain why only GCs, but not dwarf galaxies have such the abundance 
anomalies. The answer would be that the young dwarf galaxies were not compact enough for their first generation massive stars 
to sink to their central parts and merge there before exploding as supernovae or that they formed differently from GCs. 
Another more speculative scenario is that the SMSs form during
the initial monolithic collapse of the low-metallicity gas clouds from which the GCs form. 
If the gas at the centre cannot cool
as fast as it collapses then it cannot fragment and instead may form an SMS. 
Yet another possibility for the formation of SMSs assumes that a young star can simultaneously accrete material from 
the protostellar cloud through an equatorial accretion disk and lose gas (initially, with a lower rate, so that
its mass increases until the accretion and loss rates are equalized) 
in the polar directions via stellar wind \citep{tutukov:08}.  
If normal stars then form from the mass lost
from the clearly unstable SMSs then the concept from previous work of a first and second generation of GC star formation
becomes blurred. 

Any scenario explaining the origin of the abundance anomalies in GC low-mass stars by H burning 
in their more massive counterparts encounters
the mass budget problem. For the Salpeter IMF, the ratio of the mass in stars with $0.1 \msun\leq M\leq 3 \msun$ to the
total cluster mass is $M_\mathrm{low}/M_\mathrm{cl}\approx 0.76$, while the corresponding ratios for the mass lost by
AGB stars with the initial masses $3 \msun\leq M\leq 8 \msun$ and for the total mass of massive stars 
with $20 \msun\leq M\leq 120 \msun$ are 
$M_\mathrm{AGB}/M_\mathrm{cl}\approx 0.079$ and $M_\mathrm{mas}/M_\mathrm{cl}\approx 0.080$.
Observations show that a significant fraction of low-mass stars in GCs has been polluted by the products of H burning 
\citep{gratton:12}, for which the last two mass ratios would be insufficient by an order of 
magnitude. The most popular solution of this problem is to assume that (1) $M_\mathrm{cl}$ was initialy higher, at least,
by the factor of 10, (2) the pollution took place only in a central part of the young cluster, where the AGB and massive
stars rapidly migrated to, and (3) unpolluted stars from the cluster periphery were preferentially lost \citep{bekki:07}.
For the initial $M_\mathrm{cl}\sim 10^7 M_\odot$, even the more conservative estimate of \cite{gurkan:04} gives
$M_\mathrm{cc}\sim 10^4 M_\odot$. If we assume that $M_\mathrm{SMS}\sim M_\mathrm{cc}$ then,
for a typical mass of a few $10^5 \msun$ of the present-day GCs that had lost most of their unpolluted          
low-mass stars in the past and, as a result, now host comparable numbers of stars formed from the original and polluted gas,
we have $M_\mathrm{SMS}/M_\mathrm{cl}\sim 0.1$. In our solution of the mass budget problem
we first note that it takes only about $10^5$ years for our SMS models to produce the required abundance anomalies
and deposit them into the inter-stellar medium. If the original gas remained in the young cluster for a few million years,
the SMS might continue to accrete it through an equatorial disk, while losing its H-burning products via the strong
stellar wind in the polar directions. We speculate that such a state, in which the SMS has reached its maximum mass and its 
mass accretion and loss rates are equal, can persist for a few million years and, as a result, the $3\times 10^4 \msun$ star 
can partially process as much as a few $10^5 \msun$ that is comparable to the present-day cluster mass. A similar
accretion-wind model is used to explain the growth of a CO white dwarf to the Chandrasekhar mass 
in a single-degenerate scenario for SN Ia progenitors \citep{hachisu:96}. Even more relevant to our case is
the model of massive star formation with the simultaneous mass accretion from a Keplerian protostellar disc and mass ejection via magnetized
polar jets considered by \cite{seifried:12}.

The presence of IMBHs as massive as $4.7\times 10^4 M_\odot$
in GCs \citep{lutzgendorf:13} could provide, if confirmed, a strong support to our hypothesis, although                             
we cannot exclude that $M_\mathrm{IMBH}\ll M_\mathrm{SMS}$. The arguments presented by
\cite{lutzgendorf:13} are currently debated. Other works, both theoretical \citep[e.g.][]{vesperini:10} and observational
\citep{anderson:10}, either argue against the presence of
any IMBHs in GCs \citep{kirsten:12} or put $M_\mathrm{IMBH}\sim 10^3 \msol$ as an upper limit for their masses \citep{lanzoni:13,mcnamara:12}.

Given the above mentioned uncertainties and other potential problems, we regard our hypothesis as a first exploratory suggestion, 
many points of which should still be clarified and explored in detail.

\section*{Acknowledgments}

This research has been supported by the NSF
grants PHY 11-25915 and AST 11-09174 and by JINA (NSF grant PHY 08-22648). FDAH acknowledges
research support from NSERC (Canada).

\bibliography{paper}

\label{lastpage}

\end{document}